%% file: main.tex
\documentclass[conference]{IEEEtran}
\IEEEoverridecommandlockouts

\usepackage{cite}
\usepackage{amsmath,amssymb,amsfonts}
\usepackage{algorithmic}
\usepackage{graphicx}
\usepackage{xspace}
\usepackage{textcomp}
\usepackage{xcolor}
\def\BibTeX{{\rm B\kern-.05em{\sc i\kern-.025em b}\kern-.08em
    T\kern-.1667em\lower.7ex\hbox{E}\kern-.125emX}}

\input{macros}    
\begin{document}

\title{Advancing Mobile UI Testing by\\ Learning Screen Usage Semantics
}

\author{\IEEEauthorblockN{Safwat Ali Khan, skhan89@gmu.edu}
\IEEEauthorblockA{\textit{George Mason University}\\
Fairfax, VA, USA}

}

\maketitle
\begin{abstract}
\input{abstract.tex}
\end{abstract}

\begin{IEEEkeywords}
mobile application, GUI, software testing, usability, automated input generation.
\end{IEEEkeywords}
\input{introduction}
\input{research_hypothesis}

\input{expected_contribution}

\input{preliminary_work}

\input{research_plan}

\input{conclusion}

\bibliographystyle{IEEEtran}
\bibliography{ref}

\end{document}

%% file: macros.tex
\usepackage{amsmath,amssymb}
\newcommand{\aurora}{{{\sc Aurora}}\xspace}

%% file: abstract.tex
The demand for quality in mobile applications has increased greatly given users' high reliance on them for daily tasks. Developers work tirelessly to ensure that their applications are both functional and user-friendly. In pursuit of this, Automated Input Generation (AIG) tools have emerged as a promising solution for testing mobile applications by simulating user interactions and exploring app functionalities. However, these tools face significant challenges in navigating complex Graphical User Interfaces (GUIs), and developers often have trouble understanding their output. More specifically, AIG tools face difficulties in navigating out of certain screens, such as login pages and advertisements, due to a lack of contextual understanding which leads to suboptimal testing coverage. Furthermore, while AIG tools can provide interaction traces consisting of action and screen details, there is limited understanding of its coverage of higher level functionalities, such as logging in, setting alarms, or saving notes. Understanding these covered use cases are essential to ensure comprehensive test coverage of app functionalities. 

Difficulty in testing mobile UIs can lead to the design of complex interfaces, which can adversely affect users of advanced age who often face usability barriers due to small buttons, cluttered layouts, and unintuitive navigation. There exists many studies that highlight these issues, but automated solutions for improving UI accessibility needs more attention.

To address these interconnected challenges faced by mobile developers and app users, my PhD dissertation works towards advancing automated techniques for mobile UI testing. This research seeks to enhance automated UI testing techniques by learning the screen usage \textit{semantics} of mobile apps and helping them navigate more efficiently, offer more insights about tested functionalities and also improve the usability of a mobile app's interface by identifying and mitigating UI design issues. 

%% file: introduction.tex
\section{Introduction}
\label{sec:introduction}
The widespread adoption of mobile applications has changed the way people perform daily tasks, from communication and banking to healthcare and entertainment. The global mobile app market continues to expand\cite{appbrain_android_2025}, which adds extra pressure for the developers to build applications that are functional, intuitive, and user-friendly at a higher pace. Given the central role that mobile apps play in modern society, ensuring their quality is a critical concern for developers are of utmost priority to their users \cite{kong19,zeng_automated_2016}.

However, the rapid demand for mobile applications introduces significant challenges in the development lifecycle. Developers face many challenges, such as, working with evolving and fault-prone APIs~\cite{mario:fse13,Bavota:TSE15} and competing with rival apps in the marketplace, all while striving to meet user expectations following their feedbacks~\cite{palomba:icsme15}. To maintain a fast release cycle, teams often prioritize feature delivery over comprehensive testing and usability improvements \cite{kochar:ICST15}. As a result, many applications are released with complex app navigation resulting from frequent updates to the visual layouts of user interface (UI) components~\cite{fu_resistance23,older_adults_tech}. Poor usability, whether due to confusing navigation, small touch targets, or excessive visual clutter, negatively impacts user satisfaction and accessibility \cite{yu_reduceSearchSpace24}.

To streamline the testing process, researchers have developed Automated Input Generation (AIG) tools, which explore applications by simulating user interactions, such as tap, swipe, and type. These techniques can decrease manual test dependency and test time while aiming to increase test efficiency by generating a large number of interactions at a small period of time and uncover faults and does not require humans to write test cases manually~\cite{orso_software_2014,monkey_uiapplication_2022,gu_practical_2019,machiry_dynodroid_2013,sasnauskas_intent_2014,ravindranath_automatic_2014,moran_automatically_2016,baek_automated_2016,mao_sapienz_2016,dong_time-travel_2020,li_humanoid_2019,deka_rico_2017,zheng_automatic_2021,khan_aurora24}. Despite their promise, AIG tools suffer from two critical limitations: (1) their test coverage is often limited due to an inability to navigate certain UI elements, such as login screens and advertisements~\cite{wang_vet_2021,khan_aurora24}, and (2) their execution traces provide little insight into whether meaningful app use cases—such as logging in, setting an alarm, or composing a message, were successfully exercised \cite{jabbar_investigateExecutionTrace23}. This lack of interpretability makes it difficult for developers to assess how well an AIG tool has tested an application.


This research aims to address these problems with contextual understanding of screen usage semantics by utilizing learning based techniques. By understanding an app screen and its usage during test time, we help build advanced automated input generation tools, which can navigate through difficult screens. Secondly, we propose a prompt engineering technique which can automatically identify the use cases covered by a given AIG tool. This will also help developers identify overlooked areas in the application not excercised by the automated techniques. Lastly, we propose an automated technique to identify and mitigate usability issues faced by elderly population due to complex user interfaces. Together, we aim to bridge the gap between automated testing, test interpretability and usability, which can help developers build more user-friendly and robust mobile applications.   

%% file: research_hypothesis.tex
\section{Research Hypothesis}
We hypothesize that using learning-based techniques to form a semantic understanding of user interfaces and test interactions can lead to the construction of more effective testing techniques and increase the usability of mobile apps.

%% file: expected_contribution.tex
\section{Expected contribution}
Following are the contributions expected through my dissertation:
\begin{itemize}
\item Assist automated input generation (AIG) tools in exploring mobile applications by leveraging specific screen use case.
\item Interpret the actions of AIG tools based on their completed use cases.
\item Improve usability of mobile applications by identifying usability issues and redesigning UIs based on user requirements.
\end{itemize}

%% file: preliminary_work.tex
\section{Preliminary work}


In this work, we explore the advantage of semantic understanding of UI screens for the task of automated test input generation. Automated Input Generation (AIG) tools are a popular technique for mobile app testing~\cite{zeng_automated_2016,choudhary15} because of their fast paced nature at input generation. Although, these tools tend to get stuck on certain screens \cite{amalfitano}, such as login screens, advertisement screens, termed as `tarpit screens' by Wang et. al.~\cite{wang_vet_2021}, which negatively affects the test coverage. We propose a technique, \aurora ~\cite{khan_aurora24}, which leverages semantic understanding of UI screens and employs heuristics specifically aimed at navigating past the given screen type. For UI screen understanding, \aurora uses a multi-modal UI classifier, which classifies any given screen into one of 21 categories. When executed in conjunction of an AIG tool, such as APE, or Monkey, \aurora monitors the tool's activity to detect when it is stuck, and when it finds such a screen, it understands the type of screen through the multi-modal classifier. Once \aurora knows which screen it is dealing with, it executes a heuristic that is generalizable across all screens of the same category. This heuristic is designed in a way so it can exercise a specific use-case which can help the AIG tool to navigate through that particular screen.

Our findings show that \aurora significantly outperforms existing AIG tools such as APE, Monkey, and VET~\cite{wang_vet_2021}, achieving a 19.6\% increase in coverage when evaluated on 17 proprietary android apps. Analyzing the orthogonal coverage, we find that this improvement is the result of \aurora's ability to accurately classify UI screens and tailored navigation strategies, enabling screen comprehension during automated testing.

%% file: research_plan.tex
\section{Research plan}
Following up our preliminary work, we propose two more projects aligned with our expected contributions for the dissertation.  
\subsection{Automatically Labeling Use Cases Exercised by AIG Tools}

When writing test scripts for mobile apps, developers often write specific methods or code snippets to test certain functionalities of an application. In contrast, automated input generation tools do not require manually written test scripts; they generate inputs dynamically based on the application's user interface. However, these tools lack the capability to provide detailed insights into which specific app functionalities were exercised during their execution. Prior work has explored extracting lower-level information, such as UI screen descriptions\cite{xui}, screen classification \cite{khan_aurora24} or interaction information \cite{capdroid} using machine learning techniques. Regarding functionality insights, earlier studies \cite{marcus04,sitir07,poshyvanyk07,dit13} have investigated ways to localize features within source code.

We plan to address this problem by localizing use cases \cite{latte,applyingUML,umtg} from AIG tool interaction traces using learning-based techniques, including but not limited to large language models, such as GPT-4. These models can help identify the salient app functionalities exercised during the execution of these tools. 


\subsection{Automatically Adapting User Interfaces for Elderly Users}
Due to frequent user feedback, mobile app developers are required to provide regular feature updates. These updates can sometimes cause confusion as user interfaces can drastically change after a major update. Sometimes developers can make poor design choices and build cluttered user interfaces with smaller logos and fonts to fit the additional functions in a single screen. Past studies found that elderly populations struggle to cope with these changes~\cite{yu_reduceSearchSpace24,fu_resistance23,Ramdowar_mHealthElderly24,pang_evolvingPerception21}. These studies inform us about the hardships faced by elderly population while trying to use the latest technologies, and gives us an opportunity to explore the possibility of an automated technique that can help improve app usability to them. In this project, we plan to build a technique that can identify symptoms of usability issues and redesign any UI based on the requirements of elderly users.

%% file: conclusion.tex
\section{Conclusion}
In this doctoral symposium paper, I present the path to my dissertation, outlining my preliminary work, current and future projects. My research is focused on understanding mobile app usages, and exploring how learning-based techniques can help create more robust automated GUI testing techniques with interpretable actions, and improve usability via the automated identification and fixing of usability issues. My preliminary work focuses on assisting automated input generation tools in learning about the screen context to help them navigate through difficult screens. My current project aims to make automated tool interactions more interpretable from a usage perspective, so developers can understand the covered use cases reflecting the app's functionality. Lastly, I plan to work on creating an automated technique to identify and mitigate app usability issues, primarily faced by the elderly population.

\section*{Acknowledgement}
I would like to express my sincere gratitude to my research advisor, Dr. Kevin Moran, for his constant guidance and support throughout this work.